\documentclass[12pt,twocolumn]{aastex63}
\usepackage{bm}
\accepted{May 26, 2020}
\shorttitle{Magnetogenesis by a nonuniform return current}
\shortauthors{Yutaka Ohira}

\begin{document}

\title{Magnetic field generation by an inhomogeneous return current}

\author{Yutaka Ohira}
\affiliation{Department of Earth and Planetary Science, The University of Tokyo, \\
7-3-1 Hongo, Bunkyo-ku, Tokyo 113-0033, Japan}
\email{y.ohira@eps.s.u-tokyo.ac.jp}



\begin{abstract}
A new generation mechanism of the magnetic field in an inhomogeneous collisionless plasma with a beam component is proposed. 
We show that even though the current and charge neutralities are initially satisfied, the current neutrality is eventually violated if there is an inhomogeneity, so that the magnetic field is generated. 
By conducting {\it ab initio} two-dimensional particle-in-cell simulations, we demonstrate that the magnetic field is generated as expected. 
The new generation mechanism of the magnetic field can play an important role in the current universe because cosmic rays can be regarded as the beam component in the astrophysical plasma. 
We propose that the first cosmic rays generate the magnetic field with a large scale at the redshift of $z\approx20$. 
\end{abstract}



\section{Introduction}
\label{sec:1}
The magnetic field and cosmic rays (CRs) are ubiquitous in the current universe and play various roles in different environments. 
Although there are many studies about the origin of the magnetic field \citep{widrow02,subramanian19}, it is still an open problem. 
The energy density of the magnetic field is about $0.1-1\ {\rm eV}~{\rm cm}^{-3}$ in our Galaxy, which is comparable to that of CRs. 
The magnetic field of the Earth protects us from the solar wind and CRs. 
Interactions of the CRs with Earth's atmosphere and the magnetic field produce the inner Van Allen belt \citep{li17}.
Thanks to the magnetic field in galaxies, CRs can push the gas in galaxies, so that the CRs drive galactic winds \citep{jubelgas18,pakmor16,jacob18}. 
Whether the galactic winds suppress the star formation in the galaxies or not depends on the magnetic field fluctuations and the amount of CRs. 
Moreover, in clusters of galaxies, plasmas are heated by the magnetic field and CRs \citep{fujita07,fujita11}. 

In the standard picture, CRs are thought to be accelerated after the magnetic field with a sufficiently large scale is generated and amplified. 
Recently, it was shown that CRs are first accelerated at the redshift of $z\approx 20$ by supernova remnants of first stars without the large-scale magnetic field \citep{ohira19}. 
\citet{ohira19} proposed a new paradigm that the magnetic field with a small scale is generated by the Weibel instability in the first supernova remnants at $z\approx20$; the small-scale magnetic field and the supernova remnant shock accelerate the first CRs by the diffusive shock acceleration; the first CRs generate the magnetic field with a large scale while propagating to the intergalactic space. 
However, the generation mechanism of magnetic field by the first CRs was not provided. 
In this work, we provide a new generation mechanism of the large-scale magnetic field by the propagating CRs. 

A two component (electrons and ions) plasma has been often assumed in the early studies for the generation of the magnetic field. 
However, astrophysical plasmas have the propagating CRs in addition to the electron-ion plasma after the first CRs are generated. 
\citet{bell03} and \citet{miniati11} showed that the magnetic field can be generated in a plasma with a high velocity beam component if there is a nonuniform resistivity. 
In this work, we show a new generation mechanism of the magnetic field in a similar plasma with a beam but without the resistivity. 
We consider an electron-proton plasma with a beam plasma in a collisionless system. 
Since the time evolution of the magnetic field is described by the Faraday's equation, $\partial {\bm B}/\partial t = - c{\bm \nabla}\times {\bm E}$, 
understanding the electric field, ${\bm E}$, is crucial. 
If we consider a physics with a time scale longer than the electron plasma oscillation and a length scale larger than the electron inertial length, the electric field can be obtained from the generalized Ohm's law \citep{krall73}:
\begin{eqnarray}
\nonumber
&&\frac{\partial}{\partial t} \left(\sum_s q_s n_s {\bm V}_s \right)+{\bm \nabla} \cdot \left(\sum_s q_s n_s {\bm V}_s {\bm V}_s \right) \\ 
=&& \sum_s \frac{q_s^2 n_s} {m_s} \left({\bm E}+\frac{{\bm V}_s \times {\bm B}}{c} \right)  +\sum_s \frac{q_s}{m_s} \left( {\bm f}_s -{\bm \nabla} p_s \right),
\label{eq:gol}
\end{eqnarray}
where $q_s, n_s, {\bm V}_s, m_s, p_s$, and ${\bm f}_s$ are the charge, number density, velocity, mass, pressure, and external force acting on a unit volume other than the electromagnetic force. The subscript $s$ denotes the particle species.   
In this work, we use $s={\rm e,p, and\ b}$ for electrons, protons, and beam component.

If there is initially no magnetic field in a two component (electron-proton) plasma, the left-hand side and the ${\bm V}_s \times {\bm B}$ term in the generalized Ohm's law vanish. 
Then, the electric field is given by ${\bm E} = -({\bm \nabla} p_{\rm e}-{\bm f}_{\rm e})/en_{\rm e}$, 
where contributions from protons are ignored because of the large mass ratio, $m_{\rm p}/m_{\rm e}\gg1$. 
If the curl of the electric field is nonzero, the magnetic field is generated by the Faraday's equation. 
The magnetic field generation by the electron pressure term is called the Biermann battery mechanism \citep{biermann50}, 
which has been widely discussed in many astrophysical phenomena \citep{subramanian94,kulsrud97,hanayama05,shiromoto14}. 
In addition, the external force acting on electrons can generate the magnetic field.
Since photons interact with electrons strongly more than protons, 
a nonuniform photon field pushes electrons and makes the electric field, so that the magnetic field is generated \citep{uarrison70,takahashi05,durrive15}.

\section{Generation of the magnetic field}
\label{sec:2}
When a heavy beam plasma with a charge enters an electron-proton plasma, the charge and current neutralities are initially violated. 
After the time scale of the electron plasma oscillation, electrons move to neutralize the charge and current densities, so that the return current of electrons is induced and the following relations are satisfied: 
\begin{equation}
0=\sum_s q_s n_s, \ \ {\bf 0}=\sum_s q_s n_s {\bm V}_s.
\label{eq:ccn}
\end{equation}
If one of the three plasmas has some inhomogeneities, the second term on the left-hand side of Equation (\ref{eq:gol}) does not always vanish, which has not been considered for the magnetic field generation. 
To understand the effect of the new term, in this work, 
we consider the second term on the left-hand side and the electron component of the first term on the right-hand side of Equation (\ref{eq:gol}). 
Then, the electric field is given by 
\begin{equation}
{\bm E} = \frac{m_{\rm e}}{e^2n_{\rm e}}{\bm \nabla} \cdot \left(\sum_s q_s n_s {\bm V}_s {\bm V}_s \right).
\label{eq:e}
\end{equation}
We consider simple examples where the beam plasma is uniform but the electron-proton plasma has a steady-state incompressible flow before the beam plasma appears. 
After the beam plasma appears, the proton and electron densities and the proton velocity field are approximately still steady state, but the electron velocity field deviates from the steady flow because a new electron flow is induced to satisfy the current neutrality. 
For $\partial n_{\rm p}/\partial t = \partial n_{\rm e}/\partial t=0$, $\partial {\bm V}_{\rm p}/\partial t={\bf 0}$, the time evolution of the magnetic field is given by 
\begin{equation}
\frac{\partial {\bm B}}{\partial t} =  \frac{m_{\rm e}c}{e} {\bm \nabla} \times \left({\bm V}_{\rm e}  \cdot {\bm \nabla}{\bm V}_{\rm e} \right).
\label{eq:mag2}
\end{equation}
The right-hand side of the above equation can be related by the evolution of the electron vorticity, $\partial ({\bm \nabla}\times {\bm V}_{\rm e})/\partial t$. 
If the electron-proton plasma initially has some inhomogeneities, the electron return current to satisfy the current neutrality makes the vorticity of the electron plasma. 
As a result, the magnetic field and the electric current are generated. 
Next, we solve Equation (\ref{eq:mag2}) for two simple and specific examples.  
For simplicity, the charge of the beam is set to be $q_{\rm b}=e$. 
We choose the beam direction as the $x$-axis in this work, so that ${\bm V}_{\rm b}=V_{\rm b}{\bm e}_{\rm x}$, where ${\bm e}_{\rm x}$ is the unit vector in the $x$ direction.  

\subsection{Example for a nonuniform density field}
First, we consider a system where a uniform beam plasma propagates to an inhomogeneous electron-proton plasma initially at rest. 
In the proton rest frame, the charge and current neutrality conditions become 
\begin{equation}
n_{\rm e} = n_{\rm p}+n_{\rm b}, \ \ n_{\rm e}{\bm V}_{\rm e}=n_{\rm b} V_{\rm b}{\bm e}_{\rm x} .
\end{equation}
In addition, the inhomogeneous density field is assumed to be organized by entropy modes, that is, the plasma pressure is uniform. 
Then, the time evolution of the magnetic field is given by 
\begin{equation}
\frac{\partial {\bm B}}{\partial t} =  \frac{m_{\rm e}c}{2e} {\bm \nabla} \times \left(\frac{\partial V_{\rm e}^2}{\partial x}\right) {\bm e}_{\rm x} .
\label{eq:magram}
\end{equation}
The above expression means that the magnetic field is generated by the gradient of the electron ram pressure induced by the return current.

To verify Equation (\ref{eq:magram}), we perform a two-dimensional particle-in-cell (PIC) simulation with the periodic boundary condition in both directions using the public code, pCANS \citep{matsumoto13,ikeya15}. 
The PIC simulation solves the full Maxwell equations and the equation of motion for many particles, that is, the generalized Ohm's law is not assumed. 
As an example, we prepare the following plasmas as an initial condition: 
\begin{eqnarray}
n_{\rm e} &=& n_{\rm e,0} \left\{1+\delta \sin \left(\frac{2\pi}{L}(x+y)\right) \right\}^{-1/2} , \nonumber \\ \
n_{\rm p} &=& n_{\rm e} - n_{\rm b,0} ,\nonumber \\ \
n_{\rm b} &=& n_{\rm b,0},\nonumber \\ \
T_{\rm e} &=& T_{\rm 0}(n_{\rm e,0}/n_{e}) ,\nonumber \\ \
T_{\rm p} &=& T_{\rm 0}(n_{\rm e,0}/n_{e}) ,\nonumber \\ \
{\bm V}_{\rm e} &=& V_{\rm e,0}(n_{\rm e,0}/n_{e}){\bm e}_{\rm x} ,\nonumber \\ \
{\bm V}_{\rm p} &=& {\bm 0},\nonumber \\ \
{\bm V}_{\rm b} &=& V_{\rm b,0} {\bm e}_{\rm x}, \nonumber 
\end{eqnarray}
where $n_{\rm e,0}, n_{\rm b,0}, T_0, V_{\rm b,0}, V_{\rm e,0}=V_{\rm b,0}(n_{\rm b,0}/n_{\rm e,0})$, and $\delta$ are constants, and $T_{\rm e}$ and $T_{\rm p}$ are the electron and proton temperatures, and $L$ is the simulation box size. 
Assuming that ${\bm V}_{\rm e}$ is constant in time, the analytical solution to Equation (\ref{eq:magram}) is given by
\begin{equation}
{\bm B} =  \frac{2\pi^2m_{\rm e}cV_{\rm e,0}^2\delta t}{eL^2} \sin \left(\frac{2\pi}{L}(x+y)\right) \ {\bf e}_{\rm z} .
\label{eq:ana1}
\end{equation}
The magnetic field linearly increases with time. 
The increase rate becomes large for the large electron velocity, $V_{\rm e,0}=V_{\rm b,0}(n_{\rm b}/n_{\rm e})$, and the large amplitude of the fluctuation, $\delta$, but it becomes small for the large scale, $L$.
The spatially averaged energy density of the magnetic field for the z component is given by 
\begin{equation}
\frac{<B_{\rm z}^2>}{4\pi n_{\rm e,0}m_{\rm e}V_{\rm e,0}^2} = 2\pi^4\delta^2 \left(\frac{V_{\rm e,0}}{c} \right)^2 \left(\frac{L}{c/\omega_{\rm pe}} \right)^{-4}  (\omega_{\rm pe}t)^2, 
\label{eq:ana2}
\end{equation}
where the magnetic energy density is normalized by the kinetic energy density of the electron drift motion induced by the return current, and $\omega_{\rm pe}$ is the electron plasma frequency.

\begin{figure}
\epsscale{1.2}
\plotone{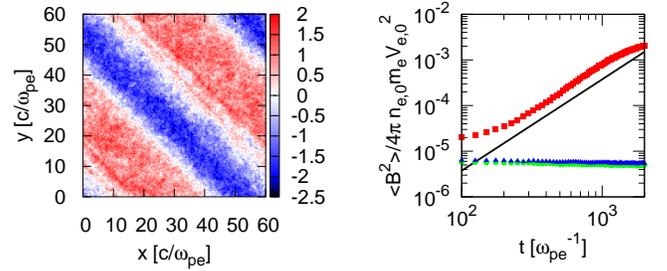}
\caption{Simulation results of the z component of the magnetic field at $t=5\times10^2\ \omega_{\rm pe}^{-1}$ (left) and the time evolution of the spatially averaged energy density of the magnetic field (right) for the first example. The magnetic field in the left panel is normalized by the amplitude of the analytical solution, Equation (\ref{eq:ana1}). The circle, triangle, and square points in the left panel show the energy density of the magnetic field for $x$, $y$, and $z$ component, respectively, and the solid line shows the analytical solution, Equation (\ref{eq:ana2}).}
\label{f1}
\end{figure}

Figure~\ref{f1} show results of the PIC simulation, where $n_{\rm b,0}/n_{\rm e,0}=0.02, V_{\rm b,0}=0.5\ c, V_{\rm e,0}=0.01\ c, \delta=0.5$, and $L=60\ c/\omega_{\rm pe}$, respectively. 
The other simulation parameters are as follows: the cell size $\Delta x = 0.02\ c/\omega_{\rm pe}$, the time step $\Delta t = 0.01\ \omega_{\rm pe}^{-1}$, the mass ratio $m_{\rm p}/m_{\rm e}=1836$, the mean thermal velocity $v_{\rm th,e} = v_{\rm th,p} (m_{\rm p}/m_{\rm e})^{1/2} =0.02\ c$, and the mean number of particle per cell for each species $n_{\rm ppc}=50$. 
The $z$ component of the magnetic field at time $t=5\times10^2\ \omega_{\rm pe}^{-1}$ is shown in the left panel of Figure~\ref{f1}. 
The right panel of Figure~\ref{f1} shows the time evolution of the spatially averaged energy density of the magnetic field. 
The circle, triangle, and square points show the simulation results for $x$, $y$, and $z$ components of the magnetic field, 
and the solid line shows the analytical solution. 
The simulation results are almost consistent with the analytical solutions. 
At around $t\sim 10^2\ \omega_{\rm pe}^{-1}$, the magnetic field fluctuation of the numerical noise dominates the physical magnetic field. 
In the later phase ($t>3\times 10^2\ \omega_{\rm pe}^{-1}$), the electron velocity, ${\bm V}_{\rm e}$, is slightly not constant, which causes the deviations from the analytical solutions.

\subsection{Example for a nonuniform velocity field}
\label{sec2b}

As the second specific example, we consider a uniform beam plasma and an electron-proton plasma which has a steady-state incompressible velocity field and a uniform density. 
The charge and current neutrality conditions become 
\begin{equation}
n_{\rm e} = n_{\rm p}+n_{\rm b}, \ \ n_{\rm e} {\bm V}_{\rm e}=n_{\rm p}{\bm V}_{\rm p}+n_{\rm b} V_{\rm b} {\bm e}_{\rm x} .
\end{equation}
Then, the time evolution of the magnetic field is given by 
\begin{equation}
\frac{\partial {\bm B}}{\partial t} =  \frac{m_{\rm e}cn_{\rm b} V_{\rm b}}{en_{\rm e}}  {\bm \nabla} \times \left(\frac{\partial {\bm V}_{\rm e}}{\partial x}\right).
\label{eq:magv}
\end{equation}
In this case, the magnetic field is generated by the advection of the electron vorticity, $(n_{\rm b}/n_{\rm e})V_{\rm b}{\bm \nabla} \times {\bm V}_{\rm e}$. 
Before the beam component appears, there is no current because electrons and protons have the same vorticity field. 
After the beam component appears, the current is generated because only the electron vorticity is advected by the flow induced by the return current. 

To verify the above equation, we perform the other PIC simulation with the periodic boundary condition in both directions. 
As an example, we prepare the following plasmas as an initial condition: 
\begin{eqnarray}
n_{\rm e} &=& n_{\rm e,0} , \nonumber \\ \
n_{\rm p} &=& n_{\rm e,0} - n_{\rm b,0} ,\nonumber \\ \
n_{\rm b} &=& n_{\rm b,0},\nonumber \\ \
{\bm V}_{\rm e} &=& V_{\rm e,0} \left \{ {\bm e}_{\rm x} + \delta \sin\left(\frac{2\pi}{L}x\right){\bm e}_{\rm z} \right\} ,\nonumber \\ \
{\bm V}_{\rm p} &=& \frac{n_{\rm e,0}-n_{\rm b,0}}{n_{\rm e,0}}V_{\rm e,0}\delta \sin\left(\frac{2\pi}{L}x\right){\bm e}_{\rm z},\nonumber \\ \
{\bm V}_{\rm b} &=& V_{\rm b,0} {\bm e}_{\rm x}. \nonumber 
\end{eqnarray}
Assuming that ${\bm V}_{\rm e}$ is constant in time, the analytical solution to Equation (\ref{eq:magv}) is given by
\begin{equation}
{\bm B} =   \frac{4\pi^2m_{\rm e}cV_{\rm e,0}^2\delta t}{eL^2}  \sin\left(\frac{2\pi}{L}x\right) \ {\bf e}_{\rm y} .
\label{eq:ana3}
\end{equation}
The magnetic field linearly increases with time as with the first example. 
The increase rate is two times larger than that of the first example.  
The spatially averaged energy density of the magnetic field is given by 
\begin{equation}
\frac{<B_{\rm y}^2>}{4\pi n_{\rm e,0}m_{\rm e}V_{\rm e,0}^2} = 8\pi^4\delta^2 \left(\frac{V_{\rm e,0}}{c} \right)^2 \left(\frac{L}{c/\omega_{\rm pe}} \right)^{-4}  (\omega_{\rm pe}t)^2.
\label{eq:ana4}
\end{equation}
%

\begin{figure}
\epsscale{1.2}
\plotone{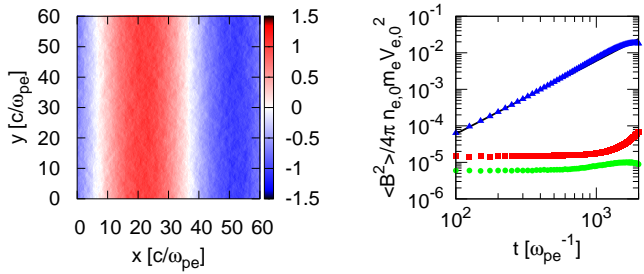}
\caption{Same as Figure~\ref{f1}, but the y component of the magnetic field (right) for the second example, where Equations. (\ref{eq:ana3}) and (\ref{eq:ana4}) are used as analytical solutions.}
\label{f2}
\end{figure}

Figure\ref{f2} show the results of the PIC simulation for the second example, where $\delta=1$ and the other parameters are the same as those of the first example. 
The $y$ component of the magnetic field at time $t=5\times10^2\ \omega_{\rm pe}^{-1}$ is shown in the left panel of Figure \ref{f2}. 
The right panel of Figure \ref{f2} shows the time evolution of the spatially averaged energy density of the magnetic field. 
All the simulation results are in very good agreement with the analytical solutions. 
Therefore, the two PIC simulations confirmed that the magnetic field is generated in a nonuniform plasma with a beam component.

\section{Discussion}
\label{sec3}
As shown in Equation (\ref{eq:mag2}), the magnetic field is generated by the electron drift velocity, $V_{\rm e}$, which compensates the beam current. If the electron drift velocity, $V_{\rm e}$, is dissipated by other microphysics before our generation mechanism works, the magnetic field cannot be generated by our mechanism but would be generated by the resistive mechanism \citep{bell03,miniati11}. Coulomb and atomic collisions, and collisionless plasma instabilities are the origin of the resistivity. For a collisional process with a mean scattering time of $t_{\rm c}$, the dissipation time, $t_{\rm dis}$, of the electron drift is estimated by 
\begin{eqnarray}
t_{\rm dis}&=&t_{\rm c} \left( \frac{L}{c/\omega_{\rm pe}} \right)^2 \nonumber \\ 
&\approx& 2\ {\rm Gyr}\left(\frac{T_{\rm e}}{1\ {\rm eV}}\right)^{3/2}\left( \frac{L}{10^{12}\ {\rm cm}} \right)^2~~,
\label{eq:dis}
\end{eqnarray}
where the Coulomb collision is assumed in the second equation. Since the dissipation time is very long, the electron drift velocity does not dissipate for sufficiently large scales even though the Coulomb collision works. The Buneman, ion acoustic, ion two stream, and ion Weibel instabilities can be considered as collisionless instabilities \citep{ohira07,ohira08,ohira19}. The Buneman instability does not occur for $V_{\rm e}<v_{\rm th,e}$. The ion acoustic and ion two stream instabilities are stable for $T_{\rm e}\approx T_{\rm p}$ or $V_{\rm e}<c_{\rm s}$, where $c_{\rm s}$ is the phase velocity of the ion acoustic mode. These stable conditions are usually satisfied because the density ratio, $n_{\rm b}/n_{\rm e}$, is usually very small, resulting in the small electron drift velocity. 
Whether the beam component is completely dissipated by the ion Weibel instability or not is still an open issue. 
Even if the beam component is dissipated, the beam components (e.g. cosmic rays) are continuously produced and propagating in the astrophysical environment. 
Therefore, the electron drift velocity and beam component are maintained during the generation of the magnetic field as long as the spatial scale is sufficiently large. 

We discuss the saturation of the magnetic field generation. 
The available current of electrons is about $q_{\rm b}n_{\rm b}V_{\rm b}\delta$ for the above two examples. 
Therefore, from the Biot-Savart law, the maximum magnetic field strength in this mechanism is $B_{\rm max}\approx 4\pi q_{\rm b}n_{\rm b}V_{\rm b}\delta L/c$.
However, the generation of the magnetic field would stop before the magnetic field strength reaches $B_{\rm max}$ 
once the beam plasma is magnetized, that is, the gyroradius of the beam plasma becomes comparable to the coherent length scale of the magnetic field, $L$. 
Furthermore, for the above two examples, the electric field starts to oscillate after $t \gtrsim L/V_{\rm e,0}$, so that the generation of the magnetic field might stop. 
Nevertheless, the magnetic field generation could continue through nonlinear feedback on the electron velocity. 
We plan to address the long-term evolution of the magnetic field generation in future work.

Recently, we discussed acceleration of first CRs at the redshift of $z\approx20 (t\approx10^8\ {\rm yr})$ \citep{ohira19}. 
In unmagnetized collisionless shocks, the Weibel instability generates small-scale magnetic fields, resulting in acceleration of the first CRs. 
The mean energy density of the first CRs is estimated to be $\sim 3\times 10^{-6}\ {\rm eV\ cm}^{-3}$ at $z\approx20$ \citep{ohira19}. 
The mean number density of the free electrons at $z\approx 20$ is $10^{-7} \ {\rm cm}^{-3}$. 
Then, the electron velocity induced by the return current is expected to be $V_{\rm e} \sim 10^4\ {\rm cm\ s}^{-1}(V_{\rm b}/c)$. 
The accretion to dark matter halos of low density baryons is almost spherically symmetrical and is stopped at the virial radius of $0.1-1\ {\rm kpc}$, while high density baryons have filamentary structures and freefall to the halo center \cite[e.g.][]{yoshida03}. Therefore, shear flows with $1\ {\rm kpc}$ scale are generated for $10^8\ {\rm yr}$. 
If we apply the second example to the large-scale structure at the early universe, 
the expected magnetic field strength at $z\approx 20$ is 
\begin{eqnarray}
B\sim 7.5\times 10^{-24}\ &{\rm G}& \left(\frac{\delta}{10^2}\right)\left(\frac{V_{\rm e,0}}{10^4\ {\rm cm\ s}^{-1}}\right)^2\nonumber \\ 
&\times&  \left(\frac{L}{1\ {\rm kpc}}\right)^{-2} \left(\frac{t}{10^{8}\ {\rm yr}}\right), 
\end{eqnarray}
where $\delta = 10^2$ is assumed because the velocity of the accretion flow is about $10^6\ {\rm cm\ s}^{-1}$ at $z\approx 20$ \citep{barkana01}. 
This is sufficiently large to be the seed of the magnetic field in the current galaxies \citep{davis99}. 
Once the large-scale magnetic field is generated, it is not dissipated even after the generation machines stop because of the long dissipation time (see equation (\ref{eq:dis})).  Therefore, at $z\approx 20$, the first CRs can generate the seed of the magnetic field in the current galaxies.

Astrophysical plasmas sometimes have a beam component in the current universe. In collisionless shocks, some upstream ions are reflected or leaking to the upstream region \citep{leroy83,spitkovsky08,tomita19}, so that the reflected or leaking ions become the beam component in the upstream rest frame. In addition to the upstream region of collisionless shocks, another beam plasma is produced in the shock downstream region if the shocks propagate to a partially ionized plasma \citep{ohira09,ohira12,ohira13,blasi12}. 
Furthermore, CRs accelerated around the shock front can be regarded as a beam plasma \citep{axford77,krymsky77,bell78,blandford78}. 
After the CRs escape from the acceleration region, the CRs can propagate far from the acceleration region. 
The escaping CRs can also be regarded as a beam plasma \citep{ohira10,ohiraetal12,fujita10}. 
The magnetic field amplification and generation around collisionless shocks are widely studied to understand the acceleration of CRs and to explain observations about high energy objects.  
Some recent particle simulations show that strong density fluctuations are generated around the shock front for high Mach number shocks \citep{caprioli14,bai15,ohira16a,ohira16b,vanmarle18,vanmarle19,tomita19}. 
The length scale of the density fluctuations is larger than or comparable to the gyroradius of ions, which is typically about $10^{10}\ {\rm cm}$. 
If we apply the first example to the upstream region of collisionless shocks, the expected magnetic field strength is 
\begin{eqnarray}
B\sim  10^{-3}\ &{\rm G}&\left(\frac{\delta}{10^0}\right) \left(\frac{n_{\rm b}/n_{\rm e}}{10^{-1}}\right)^2\nonumber \\
&\times&   \left(\frac{V_{\rm b}/c}{10^{-2}}\right)^2 \left(\frac{L}{10^{10}\ {\rm cm}}\right)^{-2} \left(\frac{t}{3\ {\rm yr}}\right), 
\end{eqnarray}
which could be larger than the magnetic field strength in the interstellar medium, $\sim 3\times 10^{-6}\ {\rm G}$. 
Therefore, the generation of the magnetic field in this work could be important for the CR acceleration. 
Since there is a finite magnetic field in the current universe, we should investigate effects of the initial magnetic field in the future. 

\section{Summary}
\label{sec4}
We have presented a new generation mechanism of the magnetic field in an inhomogeneous plasma with a beam component. 
The beam component induces the electron return current that causes the electron vorticity in the inhomogeneous plasma. 
Then, the magnetic field is generated by the electron vorticity. 
We provided analytical solutions of the magnetic field evolution for two simple examples. 
We performed two-dimensional PIC simulations, confirming that the magnetic field is generated as expected. 
This generation mechanism of the magnetic field could be important for the CR acceleration in the early and current universe. 
Moreover, at $z\approx 20$ it can generate the seed of the magnetic field of current galaxies.

\acknowledgments
The software used in this work was in part developed in pCANS at Chiba University. 
Numerical computations were carried out on the XC50 system at the Center for Computational Astrophysics (CfCA) of the National Astronomical Observatory of Japan. 
This work is supported by JSPS KAKENHI grant No. JP16K17702 and JP19H01893, and  
by Leading Initiative for Excellent Young Researchers, MEXT, Japan.


\end{document}